\documentclass[11pt]{article}

\usepackage{geometry}
\newcommand{\be}{\begin{equation}}
\newcommand{\ee}{\end{equation}}
\usepackage{amsmath}
\usepackage{amssymb}
\usepackage{latexsym}
\usepackage{epsfig}
\usepackage[vcentermath]{youngtab}

\usepackage{slashed}

\usepackage{amsmath,amssymb,amsbsy}
\usepackage{latexsym,layout,amsfonts,amssymb,fontenc,xcolor,amsthm,multirow,amsfonts,bm,textcomp}
\usepackage{hyperref}
\usepackage{empheq}
\usepackage{accents}

\newsavebox{\uuunit}
\sbox{\uuunit}
    {\setlength{\unitlength}{0.825em}
     \begin{picture}(0.6,0.7)
        \thinlines
        \put(0,0){\line(1,0){0.5}}
        \put(0.15,0){\line(0,1){0.7}}
        \put(0.35,0){\line(0,1){0.8}}
       \multiput(0.3,0.8)(-0.04,-0.02){12}{\rule{0.5pt}{0.5pt}}
     \end {picture}}

\def\2{\frac12}
\def\4{\frac14}

\def\equationautorefname~#1\null{eq.~(#1)\null
}

\hypersetup{
    bookmarks=false,         
    unicode=false,          
    pdftoolbar=true,        
    pdfmenubar=true,        
    pdffitwindow=false, 
    pdfstartview={FitH},    
    pdftitle={Exotic Dual of Type II Double Field Theory},    
    pdfauthor={Eric A.Bergshoeff, Olaf Hohm, Fabio Riccioni},     
    pdfsubject={},   
    pdfnewwindow=true,      
    colorlinks=false,       
    linkcolor=blue,          
    citecolor=blue,        
    filecolor=blue,      
    urlcolor=blue           
    linkbordercolor={blue},
    citebordercolor={purple},
    urlbordercolor={gray}
}

\begin{document}

\begin{titlepage}
\begin{center}

\hfill UG-16-71 \\

\vskip 1.5cm

{\Large \bf
Exotic Dual of Type II Double Field Theory
}

\vskip 1.5cm

{\bf  Eric A.~Bergshoeff,$^1$ Olaf Hohm,$^2$ and Fabio Riccioni$^3$}

\vskip 30pt

{\em $^1$ \hskip -.1truecm Centre for Theoretical Physics,
University of Groningen, \\ Nijenborgh 4, 9747 AG Groningen, The
Netherlands \vskip 5pt }

\vskip 15pt

{\em $^2$ \hskip -.1truecm Simons Center for Geometry and Physics, \\
Stony Brook University, \\
Stony Brook, NY 11794-3636, USA \vskip 5pt }

\vskip 15pt

{\em $^3$ \hskip -.1truecm
 INFN Sezione di Roma,   Dipartimento di Fisica, Universit\`a di Roma ``La Sapienza'',\\ Piazzale Aldo Moro 2, 00185 Roma, Italy
 \vskip 5pt }

\vskip 0.5cm

\small{e.a.bergshoeff@rug.nl, ohohm@scgp.stonybrook.edu,
Fabio.Riccioni@roma1.infn.it}

\vskip 1cm

\end{center}

\vskip 0.5cm

\begin{center} {\bf ABSTRACT}\\[3ex]
\end{center}

We perform an exotic dualization of the Ramond-Ramond fields
in type II double field theory, in which they are encoded in a Majorana-Weyl
spinor of $O(D,D)$. Starting from a first-order master action,
the dual theory in terms of a tensor-spinor of $O(D,D)$ is determined.
This tensor-spinor is subject to an exotic version of the (self-)duality
constraint needed for a democratic formulation.
We show that in components, reducing $O(D,D)$ to $GL(D)$,
one obtains the expected exotically dual theory in terms of mixed Young tableaux fields.
To this end, we generalize exotic dualizations
to self-dual fields, such as the 4-form in type IIB string theory.

\end{titlepage}

\newpage
\setcounter{page}{1} \tableofcontents


\setcounter{page}{1} \numberwithin{equation}{section}

\section{Introduction}

String theory comprises a rich spectrum of states or fields. The massless fields include
the metric, Kalb-Ramond 2-form and scalar (dilaton), together with various $p$-forms,
depending on the string theory considered, but there is also an infinite tower of massive
`higher-spin' fields, often taking values in mixed Young tableaux representations.
Even when restricting to the massless sector,
it is sometimes necessary to go beyond the minimal field content in order
to couple the various branes present in the full (non-perturbative)
string theory. For instance, in $D=10$ a 6-form needs to be introduced as the on-shell dual of the Kalb-Ramond
2-form in order to describe the NS5 brane. In recent years it has been argued from different
angles that the various dualities of string theory imply also the existence of `exotic branes' \cite{exoticbranes},
which in turn couple to fields of a more exotic nature, typically belonging to  mixed Young
tableaux representations \cite{mixedsymmetry}.

Recently, we showed how to describe, at the linearized level, such exotic dual fields 
in double field theory (DFT) \cite{Siegel:1993th,Hull:2009mi,Hohm:2010pp} 
in a T-duality or $O(D,D)$ covariant way \cite{Bergshoeff:2016ncb}.
In DFT the Kalb-Ramond field is unified
with the metric into a generalized metric ${\cal H}_{MN}$, with $O(D,D)$ indices $M,N=1,\ldots, 2D$. 
Therefore, dualizing the 2-form requires also dualizing the graviton,
which in turn leads to a mixed Young tableaux field \cite{Hull:2000zn,West:2001as}.
Moreover, additional mixed  Young tableaux fields emerge that can be interpreted as so-called
`exotic duals'
of the 2-form, implementing the  dualization procedure of \cite{Boulanger:2012df,Boulanger:2015mka}.
Remarkably, in DFT the various mixed Young tableaux representations under $GL(D)$ organize into completely
antisymmetric $O(D,D)$ tensors, including a 4-index tensor $D_{MNKL}$ for the NS sector.

In this letter, we extend the results of \cite{Bergshoeff:2016ncb} by including the Ramond-Ramond (RR)
sector of type II string theory. The difference to the NS sector is that in order to make $O(D,D)$ manifest
as a locally realized symmetry it is necessary to include for each RR $p$-form its dual $(D-p-2)$-form,
requiring a democratic formulation \cite{Fukuma:1999jt}. 
The RR fields then organize into a Majorana-Weyl spinor of $O(D,D)$,
for which a complete DFT formulation exists  \cite{Hohm:2011zr,Hohm:2011dv} 
(see \cite{Hohm:2011cp} for massive deformations and \cite{Rocen:2010bk,Thompson:2011uw} for earlier related results).
Thus, the RR fields and their conventional duals 
already enter in an $O(D,D)$ complete form, without the need to invoke
exotic dualizations. However, it is nevertheless possible to perform an exotic dualization for the RR fields,
as indeed is necessary in order to describe certain exotic branes \cite{Bergshoeff:2015cba}
and is also suggested by the Kac-Moody approach to supergravity \cite{West:2001as}.
The expected $GL(D)$ representations for the exotically dual fields can be organized into a simple
$O(D,D)$ representation, a tensor spinor $E_{MN}{}^{\alpha}$ \cite{Bergshoeff:2015cba}.
We will show here that DFT provides precisely such a formulation.

This letter is organized as follows. In sec.~2 we briefly review the exotic dualization
procedure, following \cite{Boulanger:2015mka}, and discuss the generalization to self-dual fields.
For definiteness and in order to simplify the discussion, we analyze in detail the simpler case
of a self-dual vector in $D=4$, assuming euclidean signature.
In sec.~3 we review type II DFT, and in sec.~4 we pass to an unconventional
first-order master action in order to perform the exotic dualization.
We briefly discuss how the resulting dual theory in terms of the field $E_{MN}{}^{\alpha}$
reproduces in components, breaking $O(D,D)$ to $GL(D)$, the expected result.
We close in sec.~5 with a brief summary and outlook of further exotic fields
needed in string theory.

\section{Exotic dualization of self-dual fields}

We consider here the exotic dualization of fields that are already subject to a
self-duality condition, as is the case for the 4-form in type IIB string theory or the 2-form in $(2,0)$ theories in $D=6$.
For simplicity, we analyze the case of a self-dual vector in $D=4$, which exists for euclidean
signature.

We start by reviewing the exotic dualization of the conventional Maxwell theory \cite{Boulanger:2015mka}. 
The action in terms of the field strength $F_{mn}=\partial_mA_n-\partial_nA_m$ is rewritten, up to boundary terms, as
 \be\label{BSWMaxwell}
  S \ = \ -\tfrac{1}{4}\int {\rm d}^4x\, F^{mn} F_{mn} \ = \ \int {\rm d}^4x\,\big(-\tfrac{1}{2}\partial^mA^n\partial_mA_n
  +\tfrac{1}{2}(\partial_mA^m)^2\,\big)\;,
 \ee
and then promoted to a first-order action, in terms of   fields $P_{m,n}$ and $E^{mn,k}\equiv E^{[mn],k}$, as follows:
 \be\label{Master}
  S \ = \  \int {\rm d}^4x\,\big(-\tfrac{1}{2}P^{m,n} P_{m,n}
  +\tfrac{1}{2}(P^{m}{}_{,m})^2 - E^{mn,k}\partial_{m}P_{n,k} \,\big)\;.
 \ee
The field equations for $P_{m,n}$ and $E^{mn,k}$ imply, respectively,
 \be\label{TWOequations}
 \begin{split}
  \partial^k E_{km,n} \ &= \ P_{m,n} -\eta_{mn} P^{k}{}_{,k}  \;, \\
  \partial_{[m}P_{n],k} \ &= \ 0\;.
 \end{split}
 \ee
Solving the second equation by setting $P_{m,n}=\partial_mA_n$ and re-inserting into the action,
we recover Maxwell's theory.
Equivalently, acting on the first equation with $\partial^m$ and using the `Bianchi identity'
$\partial^m\partial^k E_{km,n}=0$ we get
 \be\label{MAxwellP}
  \partial^m P_{m,n} - \partial_nP^{m}{}_{,m}  \ = \ 0\;,
 \ee
which for $P_{m,n}=\partial_mA_n$ is equivalent to the Maxwell equations.
On the other hand, solving the first equation for $P$,
\be\label{PEXPr}
 P_{m,n} \ = \ \partial^k E_{km,n} - \tfrac{1}{3}\eta_{mn}\partial^k E_{kl,}{}^{l}\;,
\ee
and back-substituting into (\ref{Master}) one obtains
a second-order action for $E$, whose field equations are obtained by inserting (\ref{PEXPr})
into the second equation of (\ref{TWOequations}).
Note that the Maxwell gauge invariance $\delta_{\lambda}A_m = \partial_m\lambda$ elevates
to a gauge invariance of the first order action given by
 \be\label{firstorderMaxwell}
  \delta_{\lambda}P_{m,n} \ = \ \partial_{m}\partial_n\lambda\;, \qquad
  \delta_{\lambda}E_{mn,k} \ = \ 2 \eta_{k[m}\partial_{n]}\lambda\;.
 \ee
There is also an extra gauge invariance associated to $E$,
 \be\label{SIGMA}
  \delta_{\Sigma}E^{mn,k} \ = \ \partial_l\Sigma^{lmn,k}\;,
 \ee
with parameter $\Sigma^{mnk,l}\equiv \Sigma^{[mnk],l}$.

We now investigate the dual theory in terms of $E$ in more detail.
Let us first decompose this field into irreducible representations as
 \be\label{DECOM}
   E^{mn,}{}_{k} \ = \ \tfrac{1}{2}\epsilon^{mnpq}
   C_{pq,k}+2\delta_{k}{}^{[m} B^{n]}\;, \qquad C_{[mn,k]} \ \equiv \ 0\;,
 \ee
where the Maxwell gauge invariance (\ref{firstorderMaxwell}) acts on the new vector $B_m$,
$\delta_{\lambda}B_m=\partial_m\lambda$.
Inserting this decomposition into (\ref{PEXPr}), one obtains
 \be\label{Pexpr}
  P_{m,n} \ = \ \partial_n B_m - \tfrac{1}{3!}\epsilon_{m}{}^{pqk} F_{pqk,n}\;, \qquad
  F_{mnk,p} \ \equiv \ 3\partial_{[m} C_{nk],p}\;.
 \ee
 The second-order field equation following from the dual action for $E$ is equivalent
 to $\partial_{[m}P_{n],k}=0$, i.e.~to
  \be
   0 \ = \  \epsilon_{mn}{}^{kl}\partial_k P_{l,p} \ = \ \partial_p\widetilde{F}_{mn}(B) +
   \partial^k F_{mnk,p}\;, \quad \widetilde{F}_{mn}(B) \ \equiv \ \tfrac{1}{2}\epsilon_{mnkl}F^{kl}(B)\;,
  \ee
where $F_{mn}(B)\equiv 2\partial_{[m}B_{n]}$.
Using the Bianchi identity $\partial^m\widetilde{F}_{mn}(B)=0$, we conclude by taking the trace that
 \be
  \partial^kF_{mnk}{}^{,m} \ = \ 0\;,
 \ee
which is the correct field equation for a $(2,1)$ field describing spin-1 in $D=4$ \cite{Boulanger:2015mka}.

We next investigate this exotic dualization for Maxwell's theory subject to a self-duality constraint, assuming  euclidean signature. Thus, the field strength satisfies
 \be\label{emduality}
  F_{mn} \ = \ \tfrac{1}{2}\epsilon_{mnkl} F^{kl}\;.
 \ee
In the first-order formulation, we then have to impose the constraint
 \be\label{Pselfduality}
  P_{[m,n]} \ = \   \tfrac{1}{2}\epsilon_{mnkl} P^{k,l}\;,
 \ee
which reduces to (\ref{emduality}) when solving the Bianchi identity for $P_{m,n}$.
Let us show that the integrability conditions of this first-order relation are compatible with
the second-order equations. To this end we act with $\partial_p$ on (\ref{Pselfduality})
and use the Bianchi identity $\partial_{[m}P_{n],k}=0$,
 \be
   \partial_pP_{m,n} - \partial_{p}P_{n,m} \ = \
   \partial_mP_{p,n} - \partial_{n}P_{p,m} \ = \ \epsilon_{mnkl}\,\partial_p P^{k,l}\;.
 \ee
Contracting this now with $\eta^{mp}$, we get
 \be
  \partial^m P_{m,n} - \partial_nP^{m}{}_{,m} \ = \ \epsilon_{mnkl}\,\partial^m P^{k,l} \ = \ 0\;,
 \ee
using again the Bianchi identity in the last step. This agrees with the second-order
equations (\ref{MAxwellP}).  It is instructive to write the (self-)duality constraint explicitly
in terms of the decomposition (\ref{DECOM}). We compute from (\ref{Pexpr}) 
 \be
  2 P_{[m,n]} =  
  -F_{mn}(B)-\tfrac{1}{2}\epsilon_{mnpq} F^{pqk}{}_{,k} \;,
 \ee
where we used the Schouten identity $0=\epsilon_{[mpqk} F^{pqk}{}_{,n]}$.  
The constraint (\ref{Pselfduality}) then implies
 \be
  F_{mn}(B) - \widetilde{F}_{mn}(B) \ = \ F_{mnp,}{}^{p}
  -\tfrac{1}{2}\epsilon_{mn}{}^{kl}\, F_{klp,}{}^{p}\;.
 \ee
Thus, the anti-self-dual part of the field strength of the vector $B_m$ is equal to the
anti-self-dual part of the trace of the `field strength' of the exotically dual field $C_{mn,k}$.
In particular, we do not obtain a first-order constraint for this field alone.
Therefore, there is no
formulation for only a (irreducible) mixed-Young-tableaux field in $D=4$ that
describes the degrees of freedom of a self-dual vector, not even on-shell.
Extra fields like the new vector $B_m$ are needed.
This can be understood by noting that for 
the gauge symmetries (\ref{SIGMA}) there is no invariant first-order field strength
for the mixed-Young-tableaux field $C_{mn,k}$, and hence there cannot be a first-order
self-duality condition. 

Let us finally note that this discussion generalizes straightforwardly 
to self-dual fields in other dimensions. For instance, for the self-dual 4-form $C_{mnkl}$ in type IIB 
string theory one promotes its derivative to a field $P_{m,klpq}$ and imposes a
Bianchi identity $\partial_{[m}P_{n],klpq}=0$ with a Lagrange multiplier field $E^{mn,klpq}$, 
which encodes the mixed Young tableaux field in the dual formulation.

\section{Ramond-Ramond fields in type II double field theory}
In this section we briefly review 
the Ramond-Ramond (RR) fields of type II double field theory, which are encoded in a Majorana-Weyl spinor
of $O(D,D)$. Our spinor conventions follow \cite{Fukuma:1999jt,Hohm:2011dv}. The Clifford algebra
 \be
  \{\Gamma_M,\Gamma_N\} \ = \ 2\eta_{MN}\;, \qquad
  \eta_{MN}  =  \begin{pmatrix}  0 & 1 \\ 1 & 0 \end{pmatrix} \;, 
 \ee
is realized in terms of fermionic oscillators $\psi_i$, $\psi^i$, with $(\psi_i)^{\dagger} = \psi^i$,
as $\Gamma_i = \sqrt{2}\psi_i$, $\Gamma^i = \sqrt{2}\psi^i$,
satisfying
 \be\label{psiClifford}
  \{\psi_i,\psi_j\} \ = \   \{\psi^i,\psi^j\} \ = \ 0\;, \qquad \{\psi_i, \psi^j\} \ = \ \delta_i{}^{j}\;.
 \ee
We define the Dirac operator with a relative factor for later convenience,
 \be
  \slashed{\partial} \ \equiv \ \tfrac{1}{\sqrt{2}}\Gamma^M\partial_M \ = \  \psi^i\partial_i +\psi_i\tilde{\partial}^i\;,
 \ee
 where $\tilde{\partial}^i$ denotes the derivative with respect to the dual coordinate.
We recall the strong constraint $\eta^{MN}\partial_M\partial_N=0$, which holds acting on arbitrary objects,
and which implies together with the Clifford algebra that  $\slashed{\partial}^2=0$.

We also need the charge conjugation matrix $C$, whose explicit expression can be found in 
\cite{Fukuma:1999jt,Hohm:2011dv}. For our purposes here it is sufficient to recall that 
 $C^{\dagger}=C^{-1}$ and 
 \be\label{Cpsirel}
  C\,\psi_i\, C^{-1} \ = \ \psi^i\;, \qquad C\,\psi^i\, C^{-1} \ = \ \psi_i\;,
 \ee
which implies for the Gamma matrices
 \be\label{CRELATIONS}
  C\,\Gamma^M\, C^{-1} \ = \ (\Gamma^{M})^{\dagger}\;, \qquad
  C^{-1}\,\Gamma^M\, C \ = \ (\Gamma^{M})^{\dagger}\;.
 \ee
The spinor representation is constructed from the Clifford vacuum $|0\rangle $ satisfying
 \be\label{CliffVacuum}
  \psi_i |0\rangle \ = \ 0 \quad \forall i\;. 
 \ee
By taking the conjugate of this equation we also conclude that $\langle 0|\psi^i = 0$ for all $i$.  
A general state is then given by
 \be
 \chi  \ = \ \sum_{p=0}^{D}\frac{1}{p!} C_{i_1\ldots i_p}\,\psi^{i_1}\cdots \psi^{i_p} |0\rangle\;, 
 \ee
which encodes the RR $p$-forms $C^{(p)}$.  
States including only even forms are of positive chirality and states including only odd forms 
are of negative chirality.  
We also use the common notation
 \begin{equation}\label{barredspinor}
\bar{\chi} \ \equiv \ \chi^\dagger  C \ = \ 
\sum_{p=0}^{D}\frac{1}{p!} C_{i_1\ldots i_p}\,\langle 0|\psi_{i_p}\cdots \psi_{i_1} C
\;.
\end{equation}

The groups Pin$(D,D)$ and Spin$(D,D)$  are the two-fold covering
groups of $O(D,D)$ and $SO(D,D)$, respectively. For a given element of the covering group
$S\in {\rm Pin}(D,D)$, there is a corresponding element $h\equiv \rho(S) \in O(D,D)$, where
$\rho:{\rm Pin}(D,D)\rightarrow O(D,D)$ is a group homomorphism,  defined implicitly by
\begin{equation}\label{SPINDef}
S \,\Gamma^M \, S^{-1}  \ = \  (h^{-1})^M{}_N\, \Gamma^N\;.
\end{equation}
Note that $+S$ and $-S$ project to the same $O(D,D)$ element $h$.
A particular ${\rm Spin}(D,D)$ element that will be useful below is ${\cal K}$,
which is the spinor representative of the generalized metric ${\cal H}^{M}{}_{N}$ with one
index raised:
 \be
  \rho({\cal K}) \ = \ {\cal H}^{\bullet}{}_{\bullet} \ = \ 
   \begin{pmatrix}  bg^{-1} & g-bg^{-1}b \\ g^{-1} & -g^{-1}b \end{pmatrix} \ \in \ O(D,D)\;, 
 \ee
where $g$ and $b$ are the metric and Kalb-Ramond 2-form.  
Denoting the spin representative of the original generalized metric ${\cal H}_{\bullet\bullet}$ by $\mathbb{S}$
and using that the charge conjugation matrix $C$ under $\rho$ actually
projects to the $O(D,D)$ metric $\eta_{MN}$ (viewed as a matrix in  $O(D,D)$),
 we have
  \be
   {\cal K} \ = \ C^{-1}\mathbb{S}\;.
  \ee
The constraints on ${\cal H}$, which read
$({\cal H}^{\bullet}{}_{\bullet})^2={\bf 1}$ and ${\cal H}_{\bullet\bullet}={\cal H}_{\bullet\bullet}^t$,
correspond to the following constraints
on $\mathbb{S}$ or equivalently ${\cal K}$,\footnote{In general dimension ${\cal K}^2=\pm{\bf 1}$, but
consistency of the self-duality constraint to be introduced below requires ${\cal K}^2={\bf 1}$. 
In the following we assume that we are in dimensions in which this is satisfied.}
 \be
  \mathbb{S}^{\dagger} \ = \ \mathbb{S}\;, \qquad {\cal K}^{2} \ = \  {\bf 1} \quad \Rightarrow \quad
  {\cal K}^{-1} \ = \  {\cal K}\;.
 \ee
We can think of $\mathbb{S}$ as being constructed from ${\cal H}$, in which case we write $\mathbb{S}=S_{\cal H}$,
but it was argued in \cite{Hohm:2011zr,Hohm:2011dv}
that a more useful perspective is to treat $\mathbb{S}$ as the fundamental field,
satisfying the above constraints.
A useful relation follows by specializing (\ref{SPINDef}) to ${\cal K}$,
 \be \label{KGammaREL}
  {\cal K}\,\Gamma^M  \ = \ {\cal H}^{M}{}_{N}\,\Gamma^N\,{\cal K}\;.
 \ee

We are now ready to define the RR action, for which we take the NS sector to be fixed,
given by a constant but otherwise arbitrary background ${\cal H}$.
The action reads
 \be\label{RRACTION}
  S_{\rm RR} \ = \ \tfrac{1}{4} \int {\rm d}^{2D} X\; (\slashed{\partial}\chi)^{\dagger}\,\mathbb{S}\,\slashed{\partial}\chi
  \ = \ \tfrac{1}{8} \int {\rm d}^{2D} X\; \partial_M\bar{\chi}\,\Gamma^M\,{\cal K}\,\Gamma^N \partial_N\chi\;,
 \ee
where the second form follows with eqs.~(\ref{CRELATIONS}) and (\ref{barredspinor}).
We have to subject the action to (self-)duality relations, since we are using a democratic formulation.
These can be written in an $O(D,D)$ covariant form as \cite{Rocen:2010bk}
 \be \label{DUalityREL}
  (1+{\cal K})\slashed{\partial}\chi \ = \ 0 \;.
 \ee
The action and duality relations are manifestly invariant under the gauge transformations
 \be\label{lambdagauge}
  \delta_{\lambda}\chi \ = \ \slashed{\partial}\lambda\;,
 \ee
due to $\slashed{\partial}^2=0$. The gauge parameter here is a Majorana-Weyl  spinor with the chirality
the opposite to that of $\chi$.

It was shown in \cite{Hohm:2011dv} how to evaluate the above action in components,
after solving the strong constraint by setting $\tilde{\partial}^i=0$, which we briefly review in the following.
To this end one has to use an explicit parametrization of the generalized metric and its spin representative,
 \be
  \mathbb{S} \ = \ S_{\cal H} \ = \ S_b^{\dagger}\, S_{g}^{-1}\,S_b\;,
 \ee
where
 \be\label{SbSgrelations}
 \begin{split}
   S_b \ &= \ e^{-\frac{1}{2}b_{ij}\psi^i\psi^j}\;,   \\
   S_{g}^{-1}\,\psi^{i_1}\cdots \psi^{i_p}|0\rangle \ &= \ \sigma\sqrt{g}\,g^{i_1j_1}\cdots g^{i_pj_p}\,
   \psi^{j_1}\cdots \psi^{j_p}|0\rangle \;,
 \end{split}
 \ee
where $\sigma=-1$ for Lorentzian signature and $\sigma=+1$ for euclidean signature.
Here we have given only the action of $S_g$ on oscillators acting on the vacuum, which is
sufficient for our purposes below.
We first observe that the naive abelian field strengths are encoded as follows,
\begin{equation}
 F \ \equiv \ \tfrac{1}{\sqrt{2}} \Gamma^M \partial_M  \chi \biggr\rvert_{\tilde{\partial}=0} \ = \
 \psi^i\partial_i\chi \quad \Rightarrow \quad F \ = \ {\rm d}C\;,
\end{equation}
using the familiar notation in which forms of different rank are combined into a single object $C$.
It is now easy to see, using eq.~(\ref{SbSgrelations}), that in the RR Lagrangian the action of $S_b$ inside
$S_{\cal H}$ changes this to the effective field strength
\begin{equation}
\widehat{F} \ = \ e^{-b_2} \wedge F\;,
\end{equation}
which is the gauge invariant field strength, given that the RR fields transform under the $b$-field
gauge symmetry.
Using again eq.~(\ref{SbSgrelations}), it is then easy to check that the RR Lagrangian reduces to
 \be
  {\cal L}_{\rm RR}\big|_{\tilde{\partial}=0} \ = \ -\tfrac{1}{4}\sqrt{g}\sum_{p=1}^{D}\frac{1}{p!}
  g^{i_1j_1}\cdots g^{i_pj_p}\widehat{F}_{i_1\ldots i_p}\widehat{F}_{j_1\ldots j_p} \;,
 \ee
which is the standard action for the RR potentials.
Similarly, it is straightforward to verify that eq.~(\ref{DUalityREL}) reduces to the conventional
duality relations for $\tilde{\partial}^i=0$.

\section{First-order action and exotic dual}

We now turn to a first-order form of the  RR action discussed in the previous section  in order to define the exotic dual.
We start from the expression (\ref{RRACTION}) and integrate by parts twice, to obtain the equivalent Lagrangian
\begin{equation}\label{partintRR}
{\cal L}_{\rm RR} \ = \
 \tfrac{1}{8}
 \; \partial_N\bar{\chi}\,\Gamma^M\,{\cal K}\,\Gamma^N \partial_M\chi\;,
\end{equation}
using that ${\cal K}$ is constant. Note that in this form the action is only gauge invariant up to
boundary terms. Next we promote $\partial_M\chi$ to an independent `vector-spinor' field $P_M$ of the same
chirality as $\chi$ and
add a Lagrange multiplier term,
 \be\label{firstorder}
  {\cal L}_{\rm 1st} \ = \ \tfrac{1}{8}\,\bar{P}_N\,\Gamma^M\,{\cal K}\, \Gamma^N P_M \ + \ \tfrac{1}{2}\,
  \partial_M\bar{P}_N \, E^{MN}\;,
 \ee
where $E^{MN}=E^{[MN]}$ is a tensor-spinor of the same chirality as $P$ for even $D$ and 
the opposite chirality for odd $D$.  
As for the second-order formulation, we have
to subject the field equations to the (self-)duality constraint, now written in terms of $P$:
\be \label{DUalityRELP}
  (1+{\cal K})\slashed{P} \ = \ 0 \;,
 \ee
where $\slashed{P}=\Gamma^MP_M$.
Varying the first-order action w.r.t.~$E^{MN}$ we obtain the constraint
 \be\label{Bianchi}
  \partial_{[M}P_{N]} \ = \ 0\;.
 \ee
This implies $P_M=\partial_M\chi$, and
 upon re-insertion into (\ref{firstorder}) and (\ref{DUalityRELP}) we recover the RR action in the form (\ref{partintRR})
 and the duality relations, respectively.
 On the other hand, varying w.r.t.~$P$ one obtains
   \be\label{PasE}
   \tfrac{1}{2}\,\Gamma^M\,{\cal K}\,\Gamma^NP_M \ = \ \partial_ME^{MN}\;,
  \ee
which are the `exotic' duality relations.
Acting with $\partial_N$ and using the Bianchi identity $\partial_M\partial_NE^{MN}=0$ we obtain
the integrability condition
 \be\label{FIRSTintegrability}
  \Gamma^M{\cal K}\, \slashed{\partial} P_M \ = \ 0\;,
 \ee
which by use of (\ref{Bianchi}), writing $P_M=\partial_M\chi$, is equivalent
to the original field equation for $\chi$.
In the following we will be interested in the theory for the exotic dual field $E_{MN}$,
obtained by eliminating $P$ using eq.~(\ref{PasE}).

Let us investigate the gauge symmetries of the first-order action corresponding to (\ref{firstorder}). First, the action is invariant,
 up to total derivatives,
 under the new gauge symmetry
  \be
   \delta_{\Sigma}E^{MN} \ = \ \partial_K\Sigma^{MNK}\;,
  \ee
 with $\Sigma^{MNK}=\Sigma^{[MNK]}$.   Second, the action is also invariant under the
 original RR gauge symmetry (\ref{lambdagauge}),
 which acts in the first-order formulation as
  \be
  \begin{split}
   \delta_{\lambda}P_M \ &= \ \partial_M\slashed{\partial}\lambda\;, \\
   \delta_{\lambda}E^{MN} \ &= \ \Gamma^{[M}\,{\cal K}\, \Gamma^{N]}\slashed{\partial}\lambda\;.
  \end{split}
 \ee
In order to prove this gauge invariance, we first consider the variation of the first-order form (\ref{firstorder})
of the RR term,\footnote{Here we used that the variation of both $P$ factors gives the same contribution,
up to total derivatives, which
can be verified in component form.}
  \be
  \begin{split}
   \delta_{\lambda}{\cal L}_{\rm RR} \ &= \ \tfrac{1}{4}\,\bar{P}_N\,\Gamma^M\,{\cal K}\,\Gamma^N
   \partial_M\slashed{\partial}\lambda \\
   \ &= \ \tfrac{1}{2}\,\bar{P}_N\,\Gamma^{[M}\,{\cal K}\,\Gamma^{N]}
   \partial_M\slashed{\partial}\lambda+\tfrac{1}{4}\,\bar{P}_N\,\Gamma^N\,{\cal K}\,\Gamma^M
   \partial_M\slashed{\partial}\lambda\\
   \ &= \  -\tfrac{1}{2}\,\partial_M\bar{P}_N\,\Gamma^{[M}\,{\cal K}\,\Gamma^{N]} \slashed{\partial}\lambda\;.
  \end{split}
 \ee
Here we used $\slashed{\partial}^2=0$ and integrated by parts with $\partial_M$ in the last step. We then
observe that the term in the last line is precisely cancelled by the variation
of $E^{MN}$ in the second term of (\ref{firstorder}), while the $\lambda$ gauge variation of $P$ in
that term drops out by the antisymmetry of $E^{MN}$.
This proves the gauge invariance of the action corresponding to (\ref{firstorder}).

Let us now return to the field equations (\ref{PasE}) in order to solve for $P$ in terms of $E$.
We first rewrite the left-hand side, using eq.~(\ref{KGammaREL}), and bring the resulting ${\cal H}$
to the other side of the equation:
 \be\label{STEP001}
  \tfrac{1}{2}\,\Gamma^M\Gamma^K\,{\cal K}\,P_M \ = \ {\cal H}^{K}{}_{N}\,\partial_ME^{MN}\;.
 \ee
Next, we contract this equation with $\Gamma_K$ and use $\Gamma_K\Gamma^M\Gamma^K
= -2(D-1)\Gamma^M$,  to obtain
 \be\label{STEP0003}
  \Gamma^M\,{\cal K} \,P_M \ = \ -\frac{1}{D-1}{\cal H}^{K}{}_{N}\,\Gamma_K\,\partial_ME^{MN}\;.
 \ee
Returning to (\ref{STEP001}) we use the Clifford algebra and compute for the left-hand side
 \be
 \begin{split}
  \tfrac{1}{2}\,\Gamma^M\Gamma^K\,{\cal K}\,P_M \ &= \ \tfrac{1}{2}\{\Gamma^M,\Gamma^K\}{\cal K}\,P_M
  -\tfrac{1}{2}\,\Gamma^K \Gamma^M\,{\cal K}\,P_M \\
  \ &= \ {\cal K}\,P^K+\frac{1}{2(D-1)}{\cal H}^{P}{}_{Q}\,\Gamma^K \Gamma_P\,\partial_LE^{LQ}\;,
 \end{split}
 \ee
where we inserted eq.~(\ref{STEP0003}) in the second line. Since this equals the right-hand side of (\ref{STEP001}),
we can solve for ${\cal K}P^M$ in terms of $E$,
 \be
  {\cal K}\,P^M \ = \ {\cal H}^{M}{}_{N}\,\partial_KE^{KN} - \frac{1}{2(D-1)}\,{\cal H}_{KL}\Gamma^M\Gamma^L
  \partial_NE^{NK}\;.
 \ee
Using ${\cal K}^2= {\bf 1}$ we can finally solve for $P^M$, obtaining the result
  \be\label{FInalPM}
  P^M \ = \  Q^M({\cal H},E)\;,
 \ee
where we defined
 \be\label{QMDEF}
  Q^M \ \equiv \  {\cal H}^{M}{}_{N}\,{\cal K}\,\partial_KE^{KN}
  - \frac{1}{2(D-1)}\,{\cal H}_{KL}\,{\cal K}\,\Gamma^M\Gamma^K
  \partial_NE^{NL}\;.
 \ee
A more compact form of this expression  is obtained by introducing the $\Sigma$ gauge invariant
`field strength'
 \be
  {\cal G}^M \ \equiv \ {\cal K}\,\partial_NE^{NM}\;,
 \ee
satisfying the Bianchi identity $\partial_M{\cal G}^M=0$.  Using eq.~(\ref{KGammaREL}) in the second term
of (\ref{QMDEF}) twice, we obtain
 \be\label{ExtraFInalPM}
  Q^M \ = \ {\cal H}^{M}{}_{N}\Big({\cal G}^N - \frac{1}{2(D-1)}\Gamma^N\,\Gamma_K\,{\cal G}^K\Big)\;.
 \ee

Back-substitution of (\ref{FInalPM}) into the Lagrangian (\ref{firstorder}) gives the second-order action for the dual field
$E^{MN}$. Its field equations are equivalent to $\partial_{[M}Q_{N]}=0$ and thus follow from
the duality relation (\ref{FInalPM}) and the Bianchi identity (\ref{Bianchi}).
Conversely, we can use the duality relation (\ref{FInalPM}) to derive the second-order
equations for the original fields. To this end, we need the Bianchi identity of the $Q^M$
defined in (\ref{QMDEF}) which reads
 \be
  \Gamma^M{\cal K}\, \slashed{\partial}\, Q_M \ \equiv \ 0\;.
 \ee
This can be verified by a direct computation, using eq.~(\ref{KGammaREL}) and the Clifford algebra
together with the Bianchi identity $\partial_M\partial_NE^{MN}=0$.
The duality relation (\ref{FInalPM}) then immediately implies the original second order equation
(\ref{FIRSTintegrability}) in terms of $P$. As usual, the duality transformations therefore swap
field equations and Bianchi identities.

We recall that the equations for the dual fields $E$ are still
subject to the first-order constraint  (\ref{DUalityRELP}), upon eliminating $P$ according to (\ref{FInalPM}),
i.e.~$(1+{\cal K})\slashed{Q}=0$.
It is instructive to verify that the integrability conditions of this (self-)duality constraint
are compatible with the second-order equations obtained from the pseudo-action, either in terms
of the original fields or the dual fields $E^{MN}$.
To this end, we act with $\partial_M$
on (\ref{DUalityRELP})  to obtain
 \be
  (1+{\cal K})\Gamma^N\partial_MP_N \ = \ 0\quad \Rightarrow \quad
  (1+{\cal K})\slashed{\partial}P_M \ = \ 0\;,
 \ee
using the Bianchi identity (\ref{Bianchi}) in the last step.
Acting with $\Gamma^M{\cal K}$ on the second equation, using ${\cal K}^2={\bf 1}$
and the Bianchi identity again, we obtain
 \be
  0 \ = \ \Gamma^M{\cal K}\,\slashed{\partial}P_M +\tfrac{1}{\sqrt{2}}\, \Gamma^M\Gamma^N\partial_NP_M
  \ = \ \Gamma^M{\cal K}\,\slashed{\partial}P_M +\tfrac{1}{\sqrt{2}}\, \partial^MP_M\;.
 \ee
Due to the Bianchi identity $P_M=\partial_M\chi$, the last term vanishes
by the strong constraint, and  indeed we recover the expected eq.~(\ref{FIRSTintegrability}).

We close this section by verifying that in components, upon solving the strong constraint
and thereby breaking $O(D,D)$ to $GL(D)$, we recover the expected exotic dualizations.
In order to simplify the presentation we will focus on a vector, subject to a self-duality constraint
in four euclidean dimensions, and match the results with those in sec.~2.
We thus assume that the fields $P_M$ and $E^{MN}$ have only the non-vanishing components
 \be\label{EPAnsatz}
  P_m \ = \ P_{m,n}\,\psi^n|0\rangle \;, \qquad
  E^{mn} \ = \ E^{mn,k}\,\psi_k \, C^{-1}|0\rangle\;, 
 \ee 
where the factor of $C$ is necessary in order for $E$ to lead to the same tensor structure
as used in sec.~2.\footnote{Equivalently, we could write $E^{mn}=E^{mn}{}_{k_1\ldots k_{D-1}}
\psi^{k_1}\ldots \psi^{k_{D-1}}|0\rangle $, in which case the term in the Lagrangian would be 
proportional to 
 \be 
  {\cal L} \ \propto \ \epsilon^{k_1\ldots k_D} E^{mn}{}_{,k_1\ldots k_{D-1}}\partial_m P_{n,k_{D}}\,, 
 \ee
c.f.~the discussion in sec.~5.1.3 in  \cite{Hohm:2011dv}. 
The definition in (\ref{EPAnsatz}) avoids the explicit epsilon tensor.}
Let us verify that $E$ has the right chirality. 
To see this note that with the `number operator' $N_F\equiv\sum_k\psi^k \psi_k$ a quick 
computation yields for the above ansatz 
 \be
  N_FP_m \ = \ P_m \quad \Rightarrow \quad 
  (-1)^{N_F}P_m \ = \ -P_m\;, 
 \ee
showing that $P_m$ has negative chirality, as it should be since it corresponds to an odd form
(1-form). Thus, in $D=4$, $E^{mn}$ should also have negative chirality and, indeed, a straightforward 
computation gives for the above ansatz $N_FE^{mn}=(D-1)E^{mn}$ and thus $(-1)^{N_F}E^{mn}=-E^{mn}$, 
as required. 
The first-order form (\ref{firstorder}) of the RR kinetic terms then reduces to
 \be
 \begin{split}
  {\cal L}_{\rm RR} \ &= \  \tfrac{1}{4}(P_{n})^{\dagger}\,C\,\psi^m\,{\cal K}\,\psi^n\, P_m
  \ = \ \tfrac{1}{4} P_{n,k} P_{m,l} \langle 0| \psi_k \, C\, \psi^m\, C^{-1} S_{\cal H}\,\psi^n\psi^l |0\rangle \\[0.5ex]
  \ &= \ \tfrac{1}{4}\sqrt{g} \, P_{n,k}\, P_{m,l}\, g^{np}\, g^{lq}\, \langle 0|\psi_k \,\psi_m\, \psi^p\, \psi^q |0\rangle\\[0,5ex]
  \ &= \ \tfrac{1}{4}\sqrt{g}\,\big( P^{m,n} P_{m,n} - (P_{n,}{}^{n})^2\big)\;,
 \end{split}
 \ee
where we used (\ref{SbSgrelations}) and that the Clifford relations (\ref{psiClifford}) and (\ref{CliffVacuum})
imply 
 \be
  \langle 0|\psi_k \,\psi_m\, \psi^p\, \psi^q |0\rangle \ = \ \delta_m{}^p \delta_k{}^q - \delta_m{}^q\delta_k{}^{p}\;.
 \ee
We infer that this reduces precisely to the $P^2$ terms in the master action (\ref{Master}),
up to an irrelevant pre-factor. Similarly, the Lagrange multiplier term in (\ref{firstorder}) reduces as 
 \be
  \tfrac{1}{2}\,\partial_M\bar{P}_N\,E^{MN} \ = \ \tfrac{1}{2}\,\partial_{m}P_{n,k} \, E^{mn,l}\,
  \langle 0|\, \psi_k\,  C\, \psi_l\, C^{-1}\, |0\rangle \ = \ \tfrac{1}{2}\,\partial_{m}P_{n,k} \, E^{mn,k}\;, 
 \ee 
where we used (\ref{Cpsirel}), giving the same term as in the Maxwell master action (\ref{Master}).
We thus recover the master action that was the starting point for the exotic dualization in sec.~2.
Moreover, the duality constraint (\ref{DUalityRELP}) yields in components the same self-duality 
constraint (\ref{Pselfduality}) as for the self-dual vector (c.f.~the discussion in sec.~5.1.3 in \cite{Hohm:2011dv}). 
We therefore have shown that  the results of this section provide the proper
$O(D,D)$ covariant exotic dualizations of the RR fields in DFT.


\section{Conclusions and Outlook}

In this letter we have applied the exotic dualization procedure of \cite{Boulanger:2015mka} to the RR fields in double field theory. This generalizes the analysis of \cite{Bergshoeff:2016ncb}, where it was shown that the dualization of the generalized metric naturally yields, together with the standard duals of the 2-form and the graviton, also the exotic dual of the 2-form. The difference between the results of \cite{Bergshoeff:2016ncb} and the analysis carried out in this letter is that in the case of the RR fields the dualization procedure is already exotic in the doubled space, while in the case of the 
generalized metric one performs a standard dualization in the doubled space, which includes the exotic dualization of the 2-form when written in components.

A natural continuation of this work would be to apply the dualization  procedure discussed in this letter
to the field $D_{MNPQ}$, which itself is the dual of the generalized metric ${\cal H}_{MN}$.
 The  dualization carried out in \cite{Bergshoeff:2016ncb} gives an action for $D_{MNPQ}$ in terms of its gauge invariant field strength. Proceeding as in this letter, one can write down a DFT action for this field in terms of the 
 gauge-dependent quantity
\begin{equation}
{\cal G}_{M, N_1 ...N_4} = \partial_M D_{N_1 ...N_4} \; ,
\end{equation}
satisfying the Bianchi identity
\begin{equation}
\partial_{[M_1} {\cal G}_{M_2 ], N_1 ...N_4} = 0 \; .
\end{equation}
In a first order formulation, the Lagrange multiplier for this constraint would be the dual  potential
$F_{M_1 M_2 , N_1 ...N_4}$. 
This field decomposes under $GL(10)$ precisely into the mixed-symmetry potentials 
given in tab.~10 of \cite{Bergshoeff:2012jb}. Such potentials can be written in a compact form as $F_{8+n , 6+m , m,n}$, where each entry denotes a set of antisymmetric indices in the mixed-symmetry representation, and $m$ and $n$ take all the possible values that are allowed by the fact that the number of indices in each set can be at most 10, with the further restriction that each set has to be greater or equal to the next. 
As expected, one of the components is the field $F_{8,6}$, which is the exotic dual of $D_6$, 
that in turn is contained in $D_{MNKL}$. 

One can also apply the dualization procedure  to the field $E_{MN}{}^{\alpha}$ discussed in this letter, 
thereby writing the DFT  action for this field in terms of
\begin{equation}
\tilde{\cal Q}_{M, N P}{}^{\alpha} = \partial_M E_{N P}{}^{\alpha} \; ,
\end{equation}
satisfying the Bianchi identity
\begin{equation}
\partial_{[M} \tilde{\cal Q}_{N], PQ }{}^{\alpha} = 0 \;. 
\end{equation}
The Lagrange multiplier in this case is a field $G_{MN, PQ}{}^{\alpha}$. 
In terms of mixed-symmetry potentials, this field decomposes as $G_{8+m,8+m ,2n,m,m}$ in the IIB case and $G_{8+m,8+m ,2n+1,m,m}$ in the IIA case. In particular, for $m=n=0$ this gives a potential $G_{8,8}$ in  the IIB case which is the exotic dual of the potential $E_8$ contained in $E_{MN}{}^\alpha$.

\vskip .7cm

\section*{Acknowledgments}
We would like to acknowledge the hospitality of the GGI (Florence) and thank the organizers of  the workshop  ``Supergravity: what next?'', where part of this work has been carried out, for creating a stimulating atmosphere. OH and FR would like to thank the University of Groningen for hospitality during the completion of this work.

\end{document}